\newcommand{\avg}[1]{\langle{#1}\rangle}
\def\avg#1{\langle{#1}\rangle}
\def\dd{_{\!_\Delta}}

\documentstyle[12pt,epsfig,JournalOfFinance]{elsart}
\begin{document} 
\begin{frontmatter} 
\title{Markovian approximation in \\ foreign exchange markets} 
\author[Aquila]{Roberto~Baviera},
\author[Roma]{Davide~Vergni}
 and
\author[Roma]{Angelo~Vulpiani\thanksref{corr}}
\address[Aquila]{Dipartimento di Fisica and INFM \\
Universit\`a dell'Aquila, I-67010 Coppito, L'Aquila, Italy}
\address[Roma]{Dipartimento di Fisica and INFM \\
Universit\`a di Roma ``La Sapienza'', I-00185 P.le A.~Moro 2, Roma, Italy}
\thanks[corr]{Corresponding author. 
Tel.: +39-06-4991 38 40;
fax:  +39-06-446 31 58;
e-mail: angelo.vulpiani@roma1.infn.it.}
\begin{abstract} 
In this paper we test the random walk hypothesis on the 
high frequency dataset 
of the bid--ask Deutschemark/US dollar
exchange rate quotes registered by the inter-bank Reuters network 
over the period October 1, 1992 to September 30, 1993. 
\\
Then we propose a stochastic model for price variation
which is able to describe some important features of the exchange market behavior.
Besides the usual correlation analysis
we have verified the validity of this model by means of 
other approaches inspired by information theory .
\\
These techniques are not only severe tests of the approximation but
also evidence some aspects of the data series 
which have a clear financial relevance. 

\smallskip


\end{abstract} 
\end{frontmatter}

\section{Introduction}

Any financial theory begins introducing 
a reasonable model for price variation in terms 
of a suitable stochastic process.
In this paper we want to
select an asset-pricing model able to describe some features
interesting from a financial point of view at {\it weak} level,
i.e. including only information arising from analysis of historical prices
and not including other pubblic (see for example \cite{Cusatis},
 \cite{Asquith}, \cite{RitterI}, \cite{RitterII}) 
or private (see \cite{Ito}) information.
\\
To attain such an aim we analyze the foreign exchange market 
which presents several advantages
compared with other financial markets.
First of all it is a very liquid market.
This feature is important because
every financial market provides a single sample path,
the set of the registered quotes of the asset.
To be sure that a statistical description makes sense, 
a reasonable requirement is that this (unique)
sequence of quotes is  
not dominated by single events or single trader's operations.
A natural candidate for such (if possible) description
is then a liquid market 
involving several billions of dollars,
daily traded by thousand of actors.
In addition the foreign exchange market has no business time limitations. 
Many market makers have branches 
worldwide so trading can occur {\it almost} continuosly.
So in the analysis one can avoid 
to consider problems involved
in the opening and closure of a particular market,
at least as a first approximation.
Finally if one considers the currency exchange, 
the returns
\begin{equation}
   r_t = \ln\frac{S_{t+1}}{S_t}\,\,
   \label{eq:return}
\end{equation}
are almost symmetrically distributed,
where $S_t$ is the price at the time $t$ 
defined as the average between bid and ask prices.
\\
In this paper we investigate the possibility to describe 
the Deutschemark/US dollar exchange (the most liquid market)
in term of a Markov process.
We consider a high frequency dataset to have statistical relevance of the 
results.
Our data, made available by Olsen and Associated,
contains all worldwide $1,472,241$ bid--ask Deutschemark/US dollar
exchange rate quotes registered by the inter-bank Reuters network 
over the period October 1, 1992 to September 30, 1993.
\\ 
One of the main problems when analyzing financial series is that
the quotes are irregulary spaced.
In section {\bf 2} we briefly describe some different 
ways to introduce time in finance,
and we discuss why we chose the {\it business} time,
i.e. the time of a transaction is the position in the sequence
of the registered quotes.
\\
The history of the efforts in the proposal of proper stochastic processes 
for price variations is very long.
An efficient foreign currency market, i.e. where prices reflect
the whole information, suggests that returns 
are independently distributed. 
Following \cite{FamaI} 
we shall call hereafter ``random walk'' a financial model where returns
are independent variables. 
Without entering into a detailed review we recall the seminal work
of \cite{Bachelier} who assumed (and tested)
that price variations follow an indipendent gaussian process.
Now it is commonly believed that returns
do not behave according to a gaussian. 
\cite{MandelbrotI} has proposed that returns are Levy-stable distributed,
still remaining independent random variables.
A recent proposal is the truncated Levy distribution model introduced by 
\cite{MantegnaStanley} which well fits financial data,
even considering them at different time lags.
\\
Because of the financial importance of correlations for 
arbitrage opportunities (see \cite{Pagan} for a review and 
\cite{BavPasSerVerVulI} for more recent results)
it is essential to introduce a non questionable technique
able to answer the question: 
can ``random walk'' models correctly describe 
return variations?
\\
In section {\bf 2} we show that ``random walk''
is inadequate to describe even qualitatively some important
features of price behavior.
We measure the probability distribution of exit time,
i.e. the lag to reach a given return amplitude $\Delta$. 
This analysis not only shows 
the presence of strong correlations but also 
leads to a natural measure of time which is intrinsic of
the market evolution, i.e. the time in which the market has such a 
fluctuation.
Following \cite{BavPasSerVerVulI} we call this time $\Delta$-trading time.
\\
In section {\bf 3}, 
measuring the time in $\Delta$-trading time,
we discuss the validity of a markovian approximation of the market behavior.
First, using the quotes, we build a Markov model of order $m$,
then, starting from the usual correlation function approach,
we consider several techniques to verify the quality
of this description.
\\
We also perform an entropic analysis, inspired by 
the \cite{Kolmogorov} $\epsilon$-entropy. 
This kind of analysis is equivalent to consider
a speculator who cares only of market fluctuation
of a given size $\Delta$ (see \cite{BavPasSerVerVulI}).
\\
Finally we use other statistical tools 
to test the validity of such an approximation,
namely the mutual information introduced by \cite{Shannon} 
and the \cite{Kullback} entropy which measures the ``discrepancy''
between the Markov
approximation and the ``true'' return process. 
\\
In section {\bf 4} we summarize and discuss the attained results.

\section{Exit times}

One of the main (and unsolved) problems in tick data analysis concerns 
the irregular spacing of quotes. There are several candidates  
to measure the time of each transaction.
\\
The first one is obviously the {\it calendar} time, i.e. the Greenwich
Meridian Time at which the transaction occurs.
The trouble with such a choice is rather evident: there are periods of no
transaction.
The simplest way (see, e.g. \cite{MantegnaStanley}) to overcome 
this difficulty is to cut ``nights'' and ``weekends'' from the signal,
i.e. assuming a zero time lag 
between the closure and reopening of the market.
Of course in a worldwide series is less evident what does it mean
``night'' or ``weekend'', but it is easy to observe that
during a day or a week there are lags when no transaction is present.
An improvement of the above procedure is to rescale the {\it calendar} time 
with a measure of market activity,
i.e. to create a new time 
scale under which in all lags the same market activity occurs.
We do not enter here in the literature on this subject,
we briefly mention the procedure in which one measures the market activity with
the average absolute number of quotes per lag (e.g. of 15 minutes) 
or with the average absolute price change over each 
lag (see \cite{Dacorogna}, \cite{MandelbrotII}).
Let us note that the weight does not change too much the time 
(roughly it is between 0.5 and 2.5), therefore there is not a big difference 
with the naive procedure.
A similar approach has been introduced by \cite{Zhang} where the time lags 
are rescaled with a properly defined instantaneous volatility.
\\
A slightly different approach is the {\it business} time.
One considers all transactions equivalent
and the time of the transaction is simply given by its position
in the sequence of quotes.
In this section we shall adopt the {\it business} time;
it looks a reasonable choice when facing a worldwide 
sequence where lags of no transaction are often a consequence of
the geografical position on the earth surface of the most important markets.
However we have to stress that, at least for some statistical features, 
there are not qualitatively differences using the {\it business} or 
the {\it calendar} time.
\\
In this section we study the distribution of the 
exit times at a given resolution $\Delta$.
This analysis will allow us to show
that the ``random walk'' models cannot be a reasonable description.
This technique will help us to understand how to analyze financial time series.
\\
Introducing
\begin{equation}
   r_{t,t_0} \equiv \ln \frac{S_t}{S_{t_0}} \,\,,
   \label{eq:rttstar}
\end{equation}
where $t_0$ is the initial {\it business} time
and $t > t_0$,
we wait until $t_1$ such as~:
\begin{equation}
   |r_{t_1,t_0}| \geq \Delta \, \, .
   \label{eq:rttbarr}
\end{equation}
Now, starting from $t_1$ with the above procedure we obtain $S_{t_2}$, and so on. 
In this way we construct the successions of
returns and exit times at given $\Delta$~:
\begin{equation}
   \{ \rho_1, \rho_2, \ldots, \rho_k, \ldots\}\:\:\:\:\:\:
   {\mathrm where} \:\:\:\:\:\: \rho_k \equiv \ln \frac{S_{t_k}}{S_{t_{k-1}}}
   \label{eq:succret}
\end{equation}
and
\begin{equation}
   \{\tau_1, \tau_2, \ldots, \tau_k, \ldots\}\:\:\:\:\:\:
   {\mathrm where} \:\:\:\:\:\: \tau_k \equiv t_k - t_{k-1}\,\,,
   \label{eq:succtau}
\end{equation}
where $|\rho_k|\geq \Delta$ by definition (see eq.(\ref{eq:rttbarr})), and
$\tau_k$ is the time after which
we have the $k$-th fluctuation of order $\Delta$ in the price.

In the following we call $k$ the $\Delta$-trading time, i.e. we enumerate
only the transactions at which a fluctuation $\Delta$ is reached.
Since the distribution of the returns is {\it almost} symmetric, 
the threshold $\Delta$ has been chosen equal for both positive and 
negative values.
\begin{figure}[htb]
 \begin{center}
  \resizebox{0.7\textwidth}{!}{\includegraphics{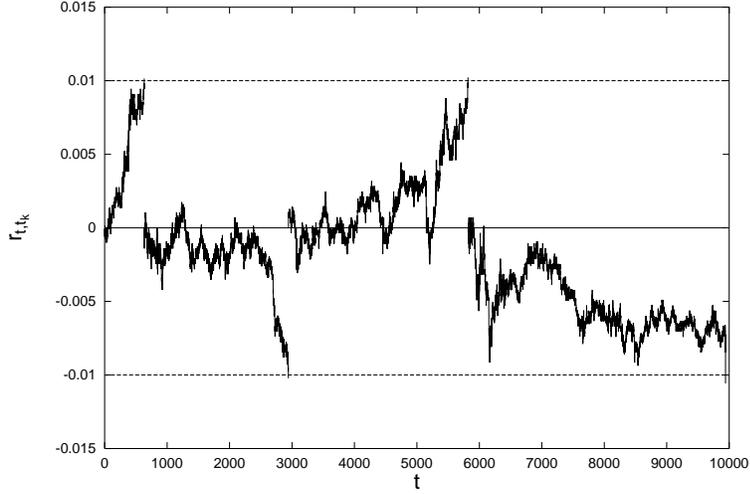}}
  \protect\caption{Evolution of $r_{t,t_k}$ with $\Delta=0.01$.
		   The $\Delta$-{\it trading} time is zero $(k=0)$
		   at the first transaction corresponding to 
		   00:00:14 of October 1, 1992 in calendar time, 
		   and $\Delta$-{\it trading} time is $4 \:\: (k=4)$
		   at 11:59:28 of October 2, 1992 (9939 {\it business} time).}
  \label{fig:distrtd}
 \end{center}
\end{figure}
\\
In the figure~\ref{fig:distrtd} we show the evolution of the returns 
$r_{t,t_k}$. 
\\
Let us now study $P\dd(\tau)$, the probability distribution function (PDF) 
of exit times (\ref{eq:succtau})
for a given size $\Delta$.
From the shape of this PDF
one can have indication if
a stochastic process can be considered a good candidate
to model price variation.
\begin{figure}[htb]
 \begin{center}
  \resizebox{0.7\textwidth}{!}{\includegraphics{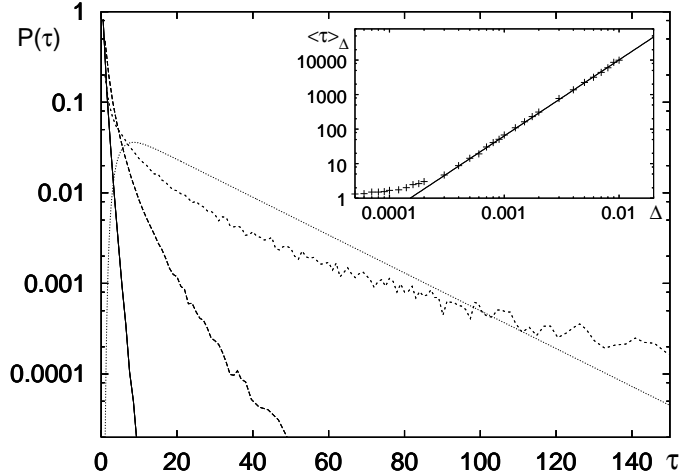}}
  \protect\caption{Probability distribution function of the exit times $\tau$ 
		   for different values of $\Delta$~:
		   $\Delta = 5\cdot 10^{-5}$ (full line),
		   $\Delta = 2\cdot 10^{-4}$ (dashed line) and 
		   $\Delta = 6\cdot 10^{-4}$ (dotted line). Those values are respectively
		   smaller, similar to and larger the typical transaction costs.
		   With the dashed-dotted line, we also show the PDF of the exit
		   times for a Wiener process (see eq. (\protect\ref{eq:wiener})) 
		   with the $\avg{\tau} = 16.85$ equal to that one obtained 
		   with $\Delta = 6\cdot 10^{-4}$. In the insert we show $\avg{\tau}$
		   versus $\Delta$. The line shows
		   the asyntotic behavior (for $\Delta > \gamma^T$)
		   of $\avg{\tau} \propto \Delta^\alpha$ with $\alpha = 2.2$.}
  \label{fig:pdfcon}
 \end{center}
\end{figure}
In figure~\ref{fig:pdfcon} we show that 
$P\dd(\tau)$ has a different shape for
$\Delta$ smaller or larger than the typical transaction
cost $\gamma^T$, where we define 
$$ \gamma_t \equiv \frac{1}{2} \ln \frac{S_t^{(ask)}}{S_t^{(bid)}} \simeq 
	      \frac{S_t^{(ask)} - S_t^{(bid)}}{2S_t^{(bid)}}\,\,$$
the transaction cost at time $t$, whose distribution has a narrow peak
around its typical value $\gamma^T=2.4\cdot~\!\!10^{-4}$.
Note that $P\dd(\tau)$ is roughly exponential 
at small $\Delta$, while 
for $\Delta$ larger than $\gamma^T$, $P\dd(\tau)$ 
clearly shows a non-exponential shape.
\\
In the insert we present $\avg{\tau}$ {\it vs.} $\Delta$.
We denote with $\avg{\cdot}$ the average of a sequence
\begin{equation}
   \avg{A} \equiv \frac{1}{L} \sum^{L}_{l=1} A_l
   \label{eq:media}
\end{equation}
where $L$ is the size of the sequence.
One has a fairly clear scaling law of the average exit time as a function
of $\Delta$ for more than three decades in $\avg{\tau}$~:
\begin{equation}
   \avg{\tau}\dd \sim \Delta^\alpha \:\:\: {\mathrm with} \:\:\: \alpha \simeq 2.2 \,\,. 
   \label{eq:tau medio}
\end{equation}
for all $\Delta$ greater than the typical transaction costs $\gamma^T$.
\\
Let us now compare the PDF from the data analysis 
with the result one obtains in ``random walk'' models.
In the case of a Wiener process in absence of drift, 
following \cite{Feller} one can find the exact solution
for the PDF~:
\begin{equation}
   P^{(W)}\dd(\tau) = \frac{4}{\pi\hat{\tau}} 
	\sum_{n=0}^\infty (-1)^n(2n+1) e^{-\frac{(2n+1)^2\tau}{\hat{\tau}}}  \,\, .
   \label{eq:wiener}
\end{equation} 
where $\hat{\tau}=(8\Delta^2)/(\pi^2\sigma^2)$ and $\sigma^2$ is the
variance of the Wiener process.
The main characteristics of this PDF are~:
\begin{itemize}
   \item it is peaked around $\avg{\tau}$.
         The similarity between average and typical value of this distribution
	 allows us to interpret the average exit time as the one we expect
	 to observe in the future with higher probability.
	 This fact is not any more true in the case of a PDF with a power-law
	 tail, and so it is more difficult to give a direct interpretation
	 of the $\avg{\tau}$ shown in the insert.
   \item It decreases exponentially at large $\tau$.
	 It is simple to understand that the exponential decay
	 is true for any ``random walk'' model.
	 In fact, in absence of correlations one has a finite probability, 
	 let us say less than $q$,
	 to exit in a finite time $\hat{\tau}$. 
	 Therefore the probability to exit after a time $\tau$
	 is less than $q^{\tau/\hat{\tau}}$, i.e. it is exponential.
	 In figure~\ref{fig:pdfcon} we also compare the normalized PDF computed
	 from financial data with the $P\dd(\tau)$ of a ``random walk''
 	 with the $\avg{\tau}$ equal to the one obtained from the signal
         with $\Delta = 6\cdot 10^{-4}$. We recall that $\gamma^T \simeq 2.4\cdot 10^{-4}$.
\end{itemize}
A model built with independent variables is unable to reproduce
$P\dd(\tau)$ and it is simple to understand that also processes 
with short memory, e.g. Markov process,
give for the exit times a PDF qualitatively similar (i.e. decaying
exponentially at large $\tau$) to that one of a Wiener process.
\\
\begin{figure}[htb]
 \begin{center}
  \resizebox{0.7\textwidth}{!}{\includegraphics{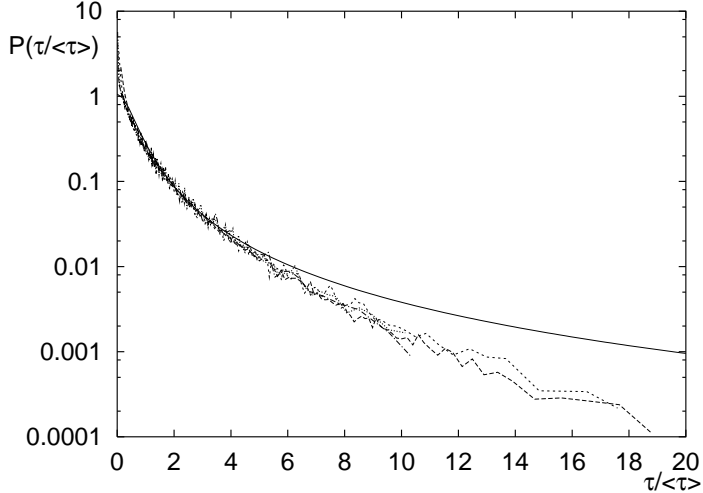}}
  \protect\caption{Data collapse of $P_\Delta(\tau/\avg{\tau})$ {\it vs}
		   $\tau/\avg{\tau}$. We have plotted four rescaled PDF~:
		   $\Delta = 4\cdot 10^{-4},\;\;\Delta = 6\cdot 10^{-4},\;\;
		    \Delta = 8\cdot 10^{-4},$ and $\Delta = 1\cdot 10^{-3}$.
		   The full curve indicates the
		   Cauchy distribution.}
  \label{fig:pdfres}
 \end{center}
\end{figure}
\\
Figure~\ref{fig:pdfres} shows the PDF data collapse,
i.e. $P\dd (\tau/\avg{\tau})\:\:{\mathit vs}\:\: \tau/\avg{\tau}$
for different $\Delta$, all greater than $\gamma^T$.

One observes that the rescaled PDF behave according to a single density function,
which is well approximated, a part for large  $\tau / \avg{\tau}$,
by the Cauchy distribution 
\begin{equation}
   P\dd \left({\frac{\tau}{\avg{\tau}}} \right) \approx 
        \frac{2a}{\pi \left({a^2 + ({\tau}/{\avg{\tau}})^2}\right)}\,\,.
   \label{eq:cauchy}
\end{equation}
The distribution $P\dd(\tau)$ shows a power-law tail up to
an exponential cut (perhaps due to the finite size of the sample analyzed).
We note that if $\Delta$ is larger than the typical transaction cost 
the probability distribution of $\tau$ has a very long tail and so 
$\avg{\tau}_\Delta$ is not the typical value of the sequence (\ref{eq:succtau}).
\\
The scaling behavior of $\avg{\tau}$ and the form of the distribution $P\dd(\tau)$ 
indicate the presence of strong correlations in the financial 
signal for $\Delta > \gamma^T$.
Furthermore the above analysis indicates the $\Delta$-trading time as 
a natural candidate to measure time. 
In the financial context this measure has the great advantage to be intrinsic of
the market evolution, i.e. it enumerates the times in which the market has such a 
fluctuation.
In the next section we shall show that, measuring the time in such a way,
a simple Markov process is a valid description of the market behavior.

\section{The markovian approximation}

In this section we construct a markovian process of order $m$ 
using a symbolic sequence obtained from the returns $\{ \rho_k \}$
at fixed $\Delta$.
We define the markovian process building 
the transition matrix with this symbolic sequence.
\\
Symbolic dynamics is a rather powerful tool to catch
the main statistical features of time series.
In order to construct a symbolic sequence from the succession $\{ \rho_k \}$
we need a {\em coarse graining} procedure to partition the range 
data and then we assign a conventional symbol to each element of the partition.
We perform the following transformation~: 
\begin{equation}
  z_k = \left \{ {\begin {array} {ccc}
		  -1 & {\mathrm if} & \rho_k < 0 \\
		  +1 & {\mathrm if} & \rho_k > 0
	  	  \end{array} } \right . \,\,\,\,.
  \label{eq:filtrononlineare}
\end{equation}
In such a way we may study a discrete stochastic process which
reproduces the feature of the original process we want to analyze.
The financial meaning of this codification is rather evident:
the symbol $-1$ occurs if the stock price decreases of percentage $\Delta$, 
while if the stock price increases of the same percentage the symbol is $1$.
In the following we indicate with $z^{(i)}$ the two possible values 
of the symbolic sequence. 
\\
Let us briefly explain the financial meaning of the return sequence 
$\{ \rho_k \}$ at fixed $\Delta$.
A speculator, who modifies his portfolio
only when a fluctuation of size $\Delta$ appears in the price sequence,
cares only of these returns.
Following \cite{BavPasSerVerVulI} we call  
{\it patient} investor such a speculator.
He performs automatically a filtering procedure~:
he rejects all the quotes which do not change 
at least of a percentage $\Delta$ of the price.
\\
Starting from this sequence we create a Markov process 
approximating the symbolic sequence.
In a Markov process of order $m$ the probability to have the symbol $z_n$ 
at the step $n$ depends only on the state of the process at the previous
$m$ steps $n-1, n-2, \ldots, n - m$.
\\
Given a sequence of $m$ symbols $Z_m=\{z^{(j_1)},z^{(j_2)},\ldots,z^{(j_m)}\}$,
we define 
$N(Z_m)$ the number of sequences $Z_m$
and $N(Z_m,j)$ the number
of times the symbol $z^{(j)}$ comes after the sequence $Z_m$.
The transition matrix of the Markov process of order $m$ is~:
\begin{equation}
   W_{Z_m,j} \equiv \frac {N(Z_m,j)}{N(Z_m)}
   \label{eq:defw}
\end{equation}
and the probability of the sequence $Z_m$ is~:
\begin{equation}
   P(Z_m) \equiv \frac {N(Z_m)}{{\mathcal{N}}_m}\,\, ,
   \label{eq:def_prob}
\end{equation}
where ${\mathcal{N}}_m = L-m$ is the total number of possible sequence
of length $m$ including superposition ($L$ is the size of the sequence).
It can be shown (see e.g. \cite{Feller}) that 
the definitions~(\ref{eq:defw}) and~(\ref{eq:def_prob}) 
are coherent in the framework of ergodic Markov processes.
\\
In the case of a Markov chain (i.e. process of order $1$) 
the transition matrix $W_{i,j}$, i.e. the probability of a transition 
in one step to the state $z^{(j)}$ 
starting from the state $z^{(i)}$, 
contains all the relevant information for the process.
Naming $N(i)$ the number of symbol $z^{(i)}$ and $N(i,j)$ the number
of symbol $z^{(j)}$ which comes after the symbol $z^{(i)}$,
the transition matrix is~:
\begin{equation}
   W_{i,j} = \frac {N(i,j)}{N(i)} \,\, .
   \label{eq:defw1}
\end{equation}
If the process is ergodic the probability $P_i$ of the state $z^{(i)}$
is given by 
\[ P_j = \sum_{i=1}^2 P_i W_{i,j} 
       = \lim_{n \rightarrow \infty} \left ({W^n} \right )_{i,j}\,\,
	 \forall i \,\, , \]
where $W^n$ is the $n$-th power of the matrix $W$.
\\
In the following we check, using various statistical approaches, if  
the markovian approximation of order $m$ 
mimics properly the price behaviour.
First of all we perform an autocorrelation analysis,
which is the most common test in financial econometrics.
Then we show that the model reproduces properly the Shannon entropy.
This quantity, as discovered by \cite{Kelly} and
shown in \cite{BavPasSerVerVulI}, has a clear financial meaning and it 
plays a central role in this field.
\\
Finally we test the validity of our approximation
using more sophisticated statistical tools of information theory.
We compare the mutual information of the signal to the one of the approximation
and we measure a ``distance'' between
the symbolic dynamics process
and the Markov approximation, with a technique based on the Kullback entropy.
\\
The results we show in the following analysis have been obtained choosing
$\Delta = 4\cdot 10^{-4}$, but they are totally independent from this
choice for $\Delta > \gamma^T$.

\subsection{Autocorrelation function}

A standard approach to verify temporal indipendence of processes
is the measure of autocorrelation function.
A first simple similarity test can be based on it~: 
if two processes have the same autocorrelation function they lose
memory of their past in a similar way.
\\
Tipically one defines {\it short memory} series if 
an exponential decay of autocorrelation function occurs, as in \cite{BrockwellDavis}.
\\
The autocorrelation function of a random process $x_t$ is  
\begin{equation}
   C(n)=\frac{\avg{x_{t+n}x_t} - \avg{x_t}^2}{\avg{x_t^2} - \avg{x_t}^2}
   \label{eq:corr}
\end{equation}
where $\avg{.}$ indicates the sequence average introduced in 
equation (\ref{eq:media}).
\\
The autocorrelation function can be easily computed for a Markov chain
described by the transition matrix $W_{i,j}$ and the probability
$P_i$ of the state $z^{(i)}$~:
\begin{equation}
   C^{(M)}(n)=\frac{\sum_{i,j}z^{(i)}z^{(j)}(W^n)_{i,j} - 
		\left( {\sum_i z^{(i)}P_i} \right)^2}
        {\sum_i z^{(i)^2} P_i - \left( {\sum_i z^{(i)}P_i} \right)^2}
   \label{eq:corrmarkov}
\end{equation}
If the Markov chain is ergodic one has
\begin{equation}
   C^{(M)}(n) \sim e^{(\ln |\lambda_2|) n} \:\:\:\: {\mathrm for\;\;large\;\;} n\,\,,
   \label{eq:appcormar}
\end{equation}
where $\lambda_2$ is the second eigenvalue of the transaction matrix
(see \cite{Feller}).
\begin{figure}[htb]
 \begin{center}
  \resizebox{0.7\textwidth}{!}{\includegraphics{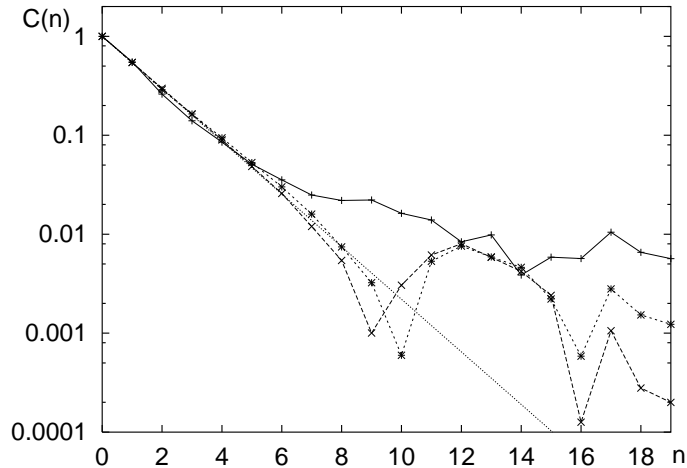}}
  \protect\caption{Autocorrelation function of return sequence $\{\rho_k\}\:(+)$,
		   Markov sequence of order $1\:\:(\times)$ and Markov sequence
		   of order $3\:\:(*)$. The line is the asymptotic analytical result
		   for a Markov process of order $1$ (see eq.(\ref{eq:appcormar})).}
  \label{fig:correlation}
 \end{center}
\end{figure}
In figure~\ref{fig:correlation} we compare the autocorrelation functions
of the return $\{ \rho_k \}$ and of the sequences of the same length
generated by Markov processes of various orders.
We show also the theoretical value for a 
Markov chain (see equation~(\ref{eq:appcormar})).
We observe there is a very good agreement between
the return sequence and the Markov process inside the statistical error.
This error, for the autocorrelation function measured 
from a finite dataset, is of the order of $O({L/\tau_c})^{-\frac{1}{2}}$
where $\tau_c$ is the correlation time 
and $L$ is the length of the sequence ($L\simeq 160000$ for $\Delta = 4\cdot10^{-4}$).
This corresponds to the value of the ``plateau'' in the figure~\ref{fig:correlation}.
\\
The same results are obtained if one computes autocorrelations
of the symbolic return sequence
$\{ z_k \}$.
This is a consequence of the fact that 
the probability density function of $\rho_k$
is peaked around the value $\pm\Delta$.

\subsection{Entropic analysis}

Let us briefly recall some basic concepts of information theory
and discuss the meaning of entropy in financial data analysis.
\\
Given a symbolic sequence $Z_n=\{z^{(j_1)},z^{(j_2)},\ldots,z^{(j_n)}\}$
of length $n$
with probability $p(Z_n)$, we define the block entropy $H_n$ as
\begin{equation}
   H_n \equiv -\sum_{Z_n} p(Z_n)\ln p(Z_n) \,\,.
   \label{def:blockentropy}
\end{equation}
The difference
\begin{equation}
   h_n \equiv H_{n + 1} - H_n
   \label{def:shannonentropy}
\end{equation}
is the average information needed to specify the symbol $z_{n+1}$ given
the previous knowledge of the sequence $\{z_1,z_2,\ldots,z_n\}$.
The series of $h_n$ is monotonically not increasing and for an {\em ergodic} 
process one has 
\begin{equation}
   h = \lim_{n \rightarrow \infty} h_n
\end{equation}
where $h$ is the \cite{Shannon} entropy.
\\
The maximum value of $h$ is $\ln(2)$ (this is because we are considering
only two-symbols sequence).
This value is reached if the process is totally uncorrelated and the 
symbols have the same probability.
We indicate, following \cite{BavPasSerVerVulI}, with {\it available} information 
the difference between the maximum entropy and its real value~: 
\[ I \equiv \ln(2) - h \;\;.\]
\cite{Khinchin} shows that if the stochastic process $\{z_1,z_2, \ldots \}$
is markovian of order $m$ then $h_n = h$ for $n \geq m$. 
We observe that the Markov process of order $m$
described by equations~(\ref{eq:defw}) and (\ref{eq:def_prob}) 
has, by definition, the same entropy $h_n$ of the symbolic sequence for all $n$
not greater than $m$. In this way we can build a markov approximation which mimics 
the originary entropy as we desire.
\begin{figure}[htb]
 \begin{center}
  \resizebox{0.7\textwidth}{!}{\includegraphics{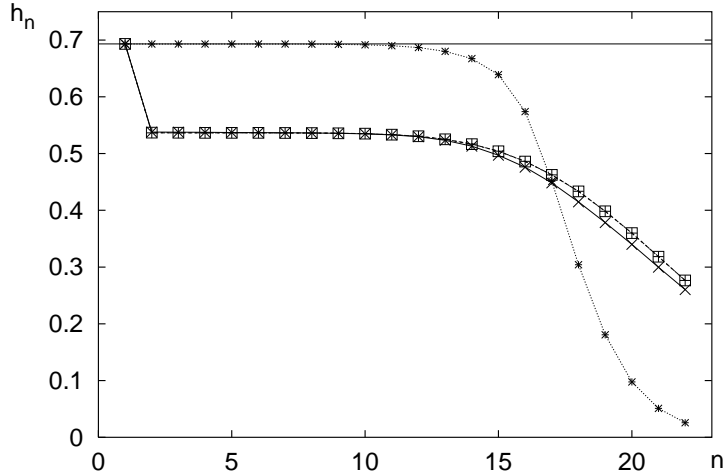}}
  \protect\caption{$h_n$ versus $n$ for the symbolic return sequence $\{z_k\}\:\:(\times)$,
		   the Markov sequence of order $1\:\:(+)$ and of order $3\:\:(\Box)$
		   (these entropies are almost indistinguishable). We also plot the $h_n$
		   for a random walk $(*)$ with its theoretical value $\log(2)$.
		   Both the Markov process and the random walk have the same number
		   of elements of the financial data with $\Delta=4\cdot 10^{-4}$.}
  \label{fig:hentr}
 \end{center}
\end{figure}
\\
From figure~\ref{fig:hentr} one observes that
the $h_n$ are consistent with those ones of a Markov process of order $1$
(the asymptotic value $h$ is reached approximatively in one step).
It is therefore natural to conjecture that such a stochastic process 
is able to mimic price variations.
Using Markov processes of order greater than one does not improve
the approximation for the entropy value in a appreciable way.
The value of $h_n$ is statistically relevant till the length
$n$ of the sequence is of the same order of $(1/h)\log(L)$
as shown in \cite{Khinchin}: this explains the folding of $h_n$ at large $n$. 
\\
Why {\it available} information is so important in the financial context?
\\
\cite{Kelly} has shown the link between {\it available} information
and the optimal
growth rate of a capital in some particular investments.
A similar idea can be applied to
the {\it patient} investor, 
i.e. a speculator who waits to modify his investment 
till a fluctuation
of size $\Delta$ is present.
He observes an {\it available} information 
different from zero.
Instead, as shown in figure~\ref{fig:hentr}, for a random walk
the {\it available} information is zero.
\\
If the returns $\{ \rho_k \}$ at fixed $\Delta$ are ruled by a Markov chain 
and if one neglets the transaction costs, \cite{BavPasSerVerVulI} prove that the 
optimal growth rate of the capital is equal to the {\it available} information.
The case with transaction costs is considered by \cite{BavieraVergni}.
\\
Inside a markovian description the {\it available} information suggests also 
the order of the process one has to consider.
It is useless to include the information coming from one further step in the past,
if one does not observe a significant increasing of the 
{\it available} information involved in the operation.
\\
Furthermore if the Markov approximation 
well describes the {\it available} information,
it is not so important for a speculator who wants to maximize
the grow rate of his capital that the markovian mimicking
is no longer good for other quantities.
However we shall show in the following that the markovian approximation
not only reproduces the {\it available} information but, in addition it is a 
proper description of the return dynamics itself.
This is important in the case one performs a more complex investment
for which a detailed model for price variation is essential 
(see \cite{Merton}).

\subsection{Mutual information}

The mutual information is a measure of the average information
one has about an event $q$ knowing the result of another event $s$.
In our case the events are the values of a process at different time.
\\
Following \cite{Shannon} we define~:
\begin{equation}
   I(n)=-\int\int P(x_t, x_{t+n})\ln\frac{P(x_t,x_{t+n})}{P(x_t)P(x_{t+n})} 
                  {\mathrm dx_t dx_{t+n}}
   \label{eq:mutual}
\end{equation}
where $P(x_t, x_{t+n})$ is the joint probability of the variables $x_t$ 
and $x_{t+n}$, and $P(x_t)$ is the probability density function of the $x_t$.\\
The main advantage of this tecnique, compared with autocorrelation function, 
is that $I(n)$ is an intrinsic property
of the process, i.e. it has the same value if we use $x_t$ or a function of it.
This because $I(n)$ depends on the probability density
function 
in such a way that the integral (\ref{eq:mutual}) is invariant under 
the change of variable $x \rightarrow y=f(x)$ .
\\
For a Markov chain the mutual information is~:
\begin{equation}
   I^{(M)}(n) = -\sum_{i,j} P_i(W^n)_{i,j} \ln\frac{P_i(W^n)_{i,j}}{P_iP_j} 
   \label{eq:mutualmarkov}
\end{equation}
and, if the Markov chain is ergodic, for large $n$ one has
\begin{equation}
   I^{(M)}(n) \sim e^{2(\ln |\lambda_2|)n} 
   \label{eq:appmutmar}
\end{equation}
where $\lambda_2$ is again the second eigenvalue of the transaction matrix.
\\
\begin{figure}[htb]
 \begin{center}
  \resizebox{0.7\textwidth}{!}{\includegraphics{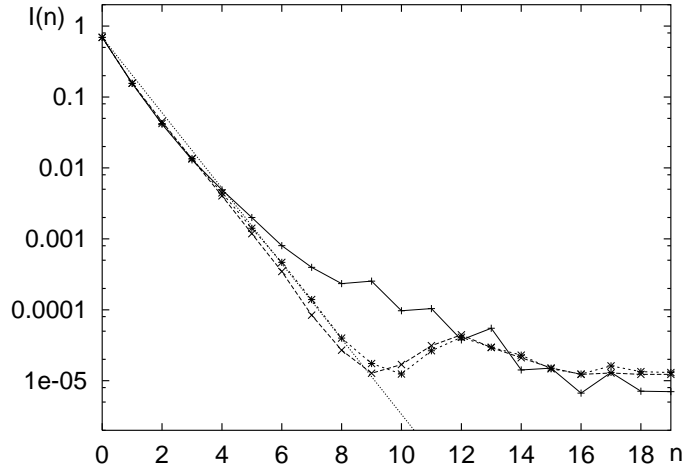}}
  \protect\caption{Mutual information of the symbolic return sequence $\{z_k\}\:\:(+)$,
		   Markov sequence of order $1\:\:(\times)$ and Markov sequence
		   of order $3\:\:(*)$. The line is the asymptotic analytical result
		   for a Markov process of order $1$ (see eq.(\ref{eq:appmutmar})).}
  \label{fig:mutual}
 \end{center}
\end{figure}
In figure~\ref{fig:mutual} we compare the mutual information
of the symbolic return series $\{z_k\}$ with the sequences 
obtained starting from Markov processes of various orders, 
and also with the expected theoretical value for a 
Markov chain (see equation~(\ref{eq:appmutmar})).
In the case of the mutual information the statistical error is of the order $L^{-1}$
(see also \cite{Roulston}) for a sequence of length $L$,
since it is computed starting from
probability distribution.
\\
Let us stress that the good agreement between the return sequence and the
Markov process for both mutual information (figure~\ref{fig:mutual}) 
and correlation function (figure~\ref{fig:correlation}) 
is a clear indicator of the validity
of the Markov approximation.

\subsection{Kullback entropy}

Given two discrete random variables $P,Q$ which can assume only $M$ different value
with probabilities  $p_i,\,q_i\;\;(i=1\ldots M)$ respectively, the Kullback
entropy of the probability distribution of the variable $P$ 
with respect to $Q$~:
\begin{equation}
   J(P|Q) \equiv \sum_{i=1}^{M} p_i\ln\frac{p_i}{q_i}
   \label{eq:kullback}
\end{equation}
is a powerful tool to measure the ``distance'' between the PDF of
those variables.
In fact, it is shown by \cite{Kullback} that
the function $J(P|Q)$ is identically zero only if the two random variables
have the same probabilities, i.e. $p_i = q_i \;\;\forall\;\;i$; otherwise $J(P|Q)>0$.
The $J(P|Q)$ is not a simmetric function of the two random variable.
To define symmetric ``distance'' between PDF,
following \cite{Kullback} we define the divergence between $\{p_i\}$ and $\{q_i\}$ as~:
\begin{equation}
   K(P|Q) \equiv J(P|Q) + J(Q|P) = \sum_{i=1}^{M} (p_i - q_i) \ln\frac{p_i}{q_i} \equiv
	    		      \sum_{i=1}^{M} (q_i - p_i) \ln\frac{q_i}{p_i} 
   \label{eq:divergence}
\end{equation}
This function is still positive definite and it is also symmetric.
\\
This is a probabilistic ``distance'' between random variables,
but we are interested to test the similarity of two stochastic processes.
To perform such a test we consider a symbolic sequence of length $n$,
$Z_n=\{z^{(j_1)},z^{(j_2)},\ldots,z^{(j_n)}\}$ 
obtained from the processes we are interested in,
and we compute the ``distance'' in the Kullback way for the PDF of such 
sequences for all $n$.
When $n \rightarrow \infty$ we test the similarity of the processes as a whole,
but, as happens for the entropy analysis, we expect for large $n$ a limitation
due to statistical errors.
\begin{figure}[htb]
 \begin{center}
  \resizebox{0.7\textwidth}{!}{\includegraphics{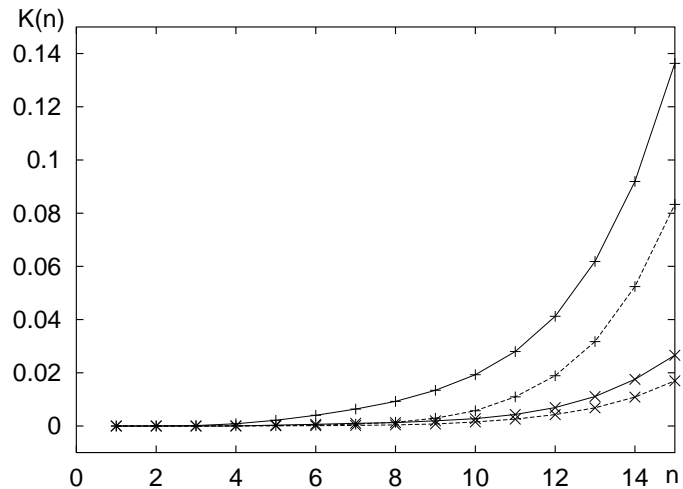}}
  \protect\caption{$K_n(P,Q)$ versus $n$. The symbols indicates the order
		   of the Markov process involved in the similarity test~:
		   $(+)$ indicate the first order and $(\times)$ indicate the third order.
		   The full line and the dashed line indicates the comparison
		   between the theoretical Markov process with the symbolic return series 
		   and the theoretical Markov process with the 
		   symbolic Markov sequence (with the same number of elements
		   of the financial data), respectively.}
  \label{fig:kullback}
 \end{center}
\end{figure}
\\
In figure~\ref{fig:kullback} we show $K_n(P|Q)$ versus $n$
where $P$ and $Q$ are respectively the Markov process and the financial process.
The sequence $Z_n$ for the financial
quotes are obtained from the symbolic return sequence $\{z_k\}$, and its 
probability is numerically computed as in equation (\ref{eq:def_prob}).
For the Markov process the probabilities of the $Z_n$ are calculated
starting from the transition matrix $W$.
In the case of a Markov chain we have~:
\[ P_n = P_{j_1}\cdot W_{j_1,j_2}\cdot W_{j_2,j_3} \cdots W_{j_{n-1},j_n} \]
We have also calculated the Kullback entropies between the PDF
of the Markov process computed theoretically 
using the transition matrix, and numerically with
a symbolic sequence (of the same length of the financial sequence).
This shows the relevance of the statistical error, and gives an indication
of the order of $n$ at which one must stop
to have statistically relevant results.
\\
All the Kullback tests are performed using Markov process of order $1$
and $3$ and is evident from figure~\ref{fig:kullback} how this
last process well reproduces the financial series.
\\
For the Kullback entropies the statistical error increases with $n$.
In order to have reasonable statistics we must restrict ourselves to take
$n \leq (1/h)\log(L)$, as in the case of entropy analysis.
\\
At the end of this section we recall that the symbolic return serie $\{z_k\}$
is chosen at fixed $\Delta$,
but the results of our analysis, i.e. the good agreement with a Markov process,
is strongly independent from the value of $\Delta$.

\section{Conclusions}

In econometric analysis it is obiously relevant to test
the validity of ``random walk'' models 
because of their strict link with market efficiency. 
In this paper  we have first discarded 
``random walk'' models as proper description of some 
features of the financial signal 
and then built a model
considering the Deutschemark/US dollar
quotes in the period from October 1, 1992 to September 30, 1993.
\\
We have developed a technique, based on the measure of exit times PDF,
which allows us to reject the random walk hypothesis.
The presence of strong correlations has been observed by means of 
an ``anomalous'' scaling of $\avg{\tau}$ and
the presence of a power law behavior of the exit times probability
density function $P\dd(\tau)$ for $\Delta > \gamma^T$.
This implies the failure of the ``random walk'' models where 
$\avg{\tau}$ scales as $\Delta^2$ and 
the $P\dd(\tau)$ tail is exponential.
\\
We want to interpret $\gamma^T$ as a natural cut-off due to the absence of 
a profitable trading rule for a {\it patient} investor 
with $\Delta$ less than $\gamma^T$~: in this case profits are less than costs.
We recall that a {\it patient} investor cares only of the quotes where 
it is present a price variation at least of a percentage $\Delta$.
\\
The main advantage of this approach is that it shows that 
this class of models gives an inadequate description 
even at the qualitative level and it suggests a new point of view 
in financial analysis.
\\
Instead of considering an arbitrary measure of time we suggest
to limit the analysis only at the times when something 
relevant from the financial point of view happens.
In particular we focus our attention on return fluctuations of size $\Delta$. 
This analysis is equivalent to the behavior of a {\it patient} speculator.
\\
We show that the returns of such a sequence 
can be approximated by a Markov model by means of several tests.
We have considered the usual autocorrelation approach obtaining 
a very good agreement inside the statistical errors.
In spite of its simplicity the autocorrelation analysis 
has the disadvantage that gives different results if one considers the 
random variable $x_t$ or a function $f(x_t)$ of it.
The mutual information is a generalization of this tool
which does not depend from the function $f$ considered.
The agreement observed for the mutual information 
is surely a severe test of similarity.
\\
A central role in the comparison between the approximation and the ``true''
signal is surely played by the {\it available} information.
Following the idea of Kelly, it has been shown by \cite{BavPasSerVerVulI} 
that this quantity corresponds, in absence of transaction costs,
to the optimal growth rate of the invested capital following a 
particular trading rule.
We show that even a Markov process of order $1$ 
mimics properly the Deutschemark/US dollar behavior.
\\
Finally we have analized a ``distance'' between processes based 
on the Kullback and Leibler entropy.
The advantage of this technique is that one 
is able to estimate the difference between the processes
with a quantity strictly connected with Shannon entropy
and then with the {\it available} information, 
i.e. the quantity of interest for a {\it patient} speculator.
\\
The Markov model we have considered in this paper not only
has the advantage to mimic very well the {\it available} information
of the financial series but also it is a good approximation
of the ``true'' process itself.
While the former is the quantity of interest for a speculator
who invests directly on the exchange market,
the latter is more interesting from both the theoretical and experimental
sides 
to have a deeper insight of the market behavior.
This asset-model will allow  
to reach a better evaluation of risk 
with the natural consequences on the derivative field.


\section*{Acknowledgment}

We would like to thank Michele Pasquini and Maurizio Serva for many
discussions on the subject.
We thank Massimo Falcioni for a critical reading of the manuscript.

\newpage
\bibliographystyle{JournalOfFinance}
\bibliography{Financial_Bibliography}

\end{document}